\documentclass[journal]{IEEEtran}
\usepackage{cite} 
\usepackage{url} 
\usepackage[utf8]{inputenc} 
\usepackage{booktabs} 
\usepackage{graphicx}
\usepackage[colorinlistoftodos]{todonotes}
\usepackage{amsmath}
\usepackage{algorithm}
\usepackage{algpseudocode}
\usepackage{textcomp}

\begin{document}
\title{DAVOS: An Autonomous Vehicle Operating System in the Vehicle Computing Era \\

{\normalsize \tt Technical Report: CAR-TR-2025-009}}


\author{
Yuxin Wang, Yuankai He, Boyang Tian, Lichen Xia, Weisong Shi\\[0.1cm]
\small Department of Computer and Information Sciences, University of Delaware\\
\small Connected and Autonomous Research Laboratory (CAR Lab)\\
\small \{yuxw, willhe, tby, lxia, weisong\}@udel.edu
}

\maketitle

\begin{abstract}
Vehicle computing represents a fundamental shift in how autonomous vehicles are designed and deployed, transforming them from isolated transportation systems into mobile computing platforms that support both safety-critical, real-time driving and data-centric services. In this setting, vehicles simultaneously support real-time driving pipelines and a growing set of data-driven applications, placing increased responsibility on the vehicle operating system to coordinate computation, data movement, storage, and access. These demands highlight recurring system considerations related to predictable execution, data and execution protection, efficient handling of high-rate sensor data, and long-term system evolvability, commonly summarized as Safety, Security, Efficiency, and Extensibility (SSEE). Existing vehicle operating systems and runtimes address these concerns in isolation, resulting in fragmented software stacks that limit coordination between autonomy workloads and vehicle data services.

This paper presents DAVOS, the Dependable Autonomous Vehicle Operating System, a unified vehicle operating system architecture designed for the vehicle computing context. DAVOS integrates Sensor-In-Memory communication to support low-latency and bounded sensor-to-application data paths, real-time scheduling to coordinate time-sensitive workloads, and a Context-aware Risk Index runtime that provides safety awareness during driving operation. To support data-centric vehicle services, DAVOS incorporates Autonomous Vehicle Storage for hierarchical and queryable data management, Privacy-aware Confidential Computing for controlled and privacy-preserving data access, and a Vehicle Programming Interface that exposes standardized abstractions across hardware, data, computation, and services. Together, DAVOS provides a cohesive operating system foundation that supports both real-time autonomy and extensible data-driven service within a single system framework.



\end{abstract}

\section{Introduction}
\label{section1:intro}

With the rapid integration of high-performance onboard computing platforms, advanced sensors, and pervasive connectivity, modern vehicles are evolving beyond traditional transportation systems into mobile computing platforms. This transformation, often described as vehicle computing, enables vehicles to continuously sense, process, store, and exchange large volumes of data, supporting a wide range of edge-enabled services alongside driving functions. In this paradigm, a vehicle effectively operates as a data-driven computer on wheels, where computation, storage, and communication are integral to system functionality rather than auxiliary components~\cite{lu2023vehicle}. 

Autonomous vehicles (AV) exemplify this shift toward vehicle computing. Equipped with heterogeneous compute resources and rich multimodal sensors, an autonomous vehicle simultaneously fulfills multiple roles. First, it operates as a real-time transportation system, executing perception, planning, and control pipelines under strict timing and safety constraints to ensure safe vehicle operation~\cite{bathla2022autonomous}. Second, it increasingly supports data-centric and service-oriented workloads, including in vehicle infotainment, accident and fault analysis, predictive maintenance, and fleet-level analytics. These emerging applications rely on persistent access to historical and real-time vehicle data, placing new demands on onboard storage, data management, and system-level coordination~\cite {liu2020computing, liu2019edge}. 

As vehicle functionality expands, the operating system becomes a central component in managing these diverse workloads. A vehicle operating system (OS) can be viewed as a unified development and execution platform that spans multiple vehicle domains, coordinating computation, communication, and data across hardware and software boundaries~\cite{VectorVehicleOS2022}. In practice, however, existing vehicle operating systems and runtimes are developed in a fragmented manner. Commercial platforms such as Ford SYNC~\cite{fordSYNC} and Android Automotive OS~\cite{androidAutomotiveOS} primarily target infotainment, human-machine interaction, and application ecosystems, and provide limited support for deterministic execution required by autonomous driving pipelines. In contrast, safety-critical driving functions are commonly deployed on dedicated real-time operating systems such as QNX~\cite{qnxRTOS}, resulting in separate and isolated software stacks within the same vehicle.

This separation leads to autonomous workloads and data-driven services coexisting through loosely coupled interfaces rather than sharing a common system substrate. As a result, resource management, data movement, and scheduling decisions are made independently across stacks, limiting system-level efficiency, extensibility, and coordination. These limitations motivate the need for an autonomous vehicle operating system designed explicitly for the vehicle computing era. Such a system should support predictable and safe execution for real-time driving workloads while also enabling extensible and secure data-driven services, bringing communication, computation, storage, and scheduling together within a unified system architecture.

This paper presents \textbf{DAVOS} (Dependable Autonomous Vehicle Operating System), an integrated operating system stack that rethinks OS design for the autonomous vehicle in the vehicle computing era. DAVOS supports both real-time autonomous driving and the broader vehicle service ecosystem through a unified architecture. Following the data flow from hardware to application, from real-time autonomy to data-driven applications, DAVOS includes:
\begin{itemize}
    \item \textbf{Sensor-In-Memory Communication (SIM)}, which provides deterministic and efficient data transfer between sensors and algorithms.
    \item \textbf{Real-Time Scheduling}, which maintains predictable execution and confidence-driven coordination across autonomy pipelines.
    \item \textbf{Context-aware Risk Index (CRI)}, which evaluates directional and contextual risk and guides the safety-critical control behavior.
    \item \textbf{Autonomous Vehicle Storage (AVS)}, a hierarchical and queryable storage system that supports real-time append logging, retrieval, and long-term data retention and transfer.
    \item \textbf{Privacy-aware Confidential Computing (PaCC)}, which enables purpose-bound third-party data access without exposing sensitive information.
    \item A \textbf{Vehicle Programming Interface (VPI)}, which provides standardized and portable abstractions across hardware, data, computation, services, and system management.
\end{itemize}

The rest of this paper is organized as follows. Section~\ref{section2:bg} reviews the background of Autonomous Driving and Vehicle computing, its unique requirements, and the limitations of existing vehicle operating systems. Section~\ref{section3:system} introduces the system architecture of DAVOS. Sections~\ref{section4:sim} to~\ref{section8:vpi} present each subsystem in detail: SIM, Real-time Scheduling, CRI, AVS, PaCC, and VPI. Finally, Section~\ref{section10:conclude} concludes with open challenges and outlines a roadmap for the future of autonomous vehicle operating systems.

\section{Motivation}
\label{section2:bg}
\begin{figure*}[!ht]
\centering
\includegraphics[width=\linewidth]{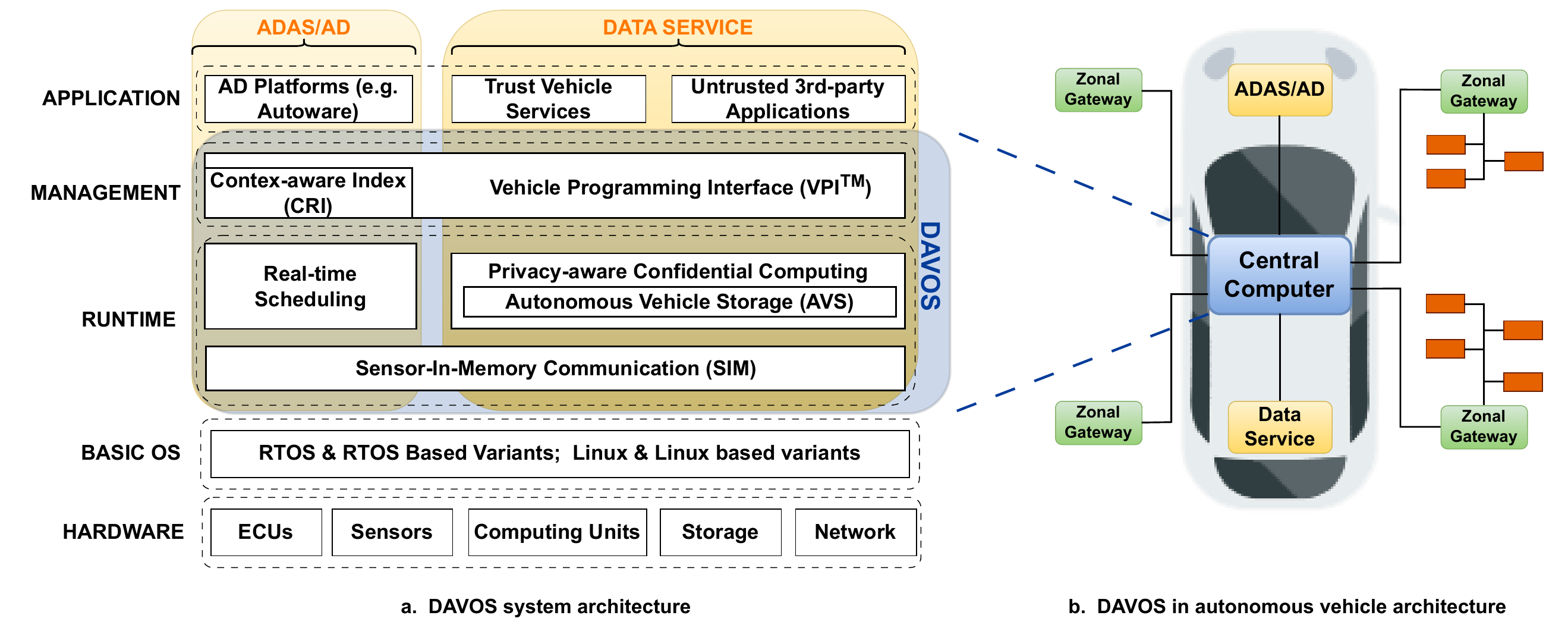} 
\caption{DAVOS architecture and vehicle deployment. In the figure, ADAS stands for Advanced Driver Assistance Systems, AD denotes Autonomous Driving, RTOS stands for Real-Time Operating System, and ECUs represent Electronic Control Units. (a) DAVOS internal architecture, illustrating the data flow from hardware to applications and the separation between real-time ADAS/AD workloads and data-centric vehicle services. 
(b) Deployment of DAVOS on the vehicle central computer, interfacing with zonal gateways to support both driving and data services. }
\label{fig:system_arch}
\end{figure*}

\subsection{Autonomous Driving and Vehicle Computing}
Autonomous vehicles are cyber-physical systems designed to perceive their environment, make driving decisions, and control vehicle motion with limited or no human intervention~\cite{bathla2022autonomous}. To achieve this functionality, an autonomous vehicle continuously executes perception, planning, and control pipelines that transform raw sensor inputs into actuation commands. These pipelines operate within closed feedback loops and interact directly with the physical world, where execution delays or timing variability can affect vehicle behavior~\cite{shalev2017formal, wang2020safety}. This execution model places emphasis on predictable task scheduling, timely data delivery, and isolation among driving-related workloads, motivating operating system support for real-time execution and safety-aware coordination.

Beyond autonomous driving, vehicles are increasingly viewed as mobile computing platforms that host a broader set of computation and data services, a paradigm often referred to as vehicle computing. In this context, the vehicle not only executes real-time control workloads, but also ingests, stores, and analyzes large volumes of heterogeneous data generated by onboard sensors and systems. Emerging data-centric services and applications such as infotainment, accident and fault analysis, predictive maintenance, and fleet-level analytics rely on persistent data access and flexible computation over historical and real-time vehicle data. Supporting these workloads introduces additional operating system considerations, including efficient storage management, and privacy-preserved data sharing~\cite{kato2018autoware,lu2023vehicle}.

Together, these technological and architectural changes motivate four recurring considerations for vehicle operating systems. \textbf{S}afety relates to supporting predictable execution and runtime safety awareness for driving workloads with stringent timing sensitivity. \textbf{S}ecurity concerns protecting onboard data and software execution as vehicles store sensitive information and expose interfaces to external services. \textbf{E}fficiency emphasizes low-latency handling of high-rate multimodal data, including timely sensor data delivery and resource-aware data management. \textbf{E}xtensibility reflects the expectation that vehicle software platforms will evolve over time, supporting new applications and services through stable and consistent programming abstractions. Collectively, these four principles—\textbf{Safe}, \textbf{Secure}, \textbf{Efficient}, and \textbf{Extendible} (\textbf{SSEE})—provide a structured lens for reasoning about operating system support in autonomous and connected vehicle computing.

\subsection{Limitations of Existing Vehicle Operating Systems}
Current vehicle software stacks are built on a diverse set of operating systems and runtimes, each optimized for a specific domain, resulting in fragmented system support for autonomous and data-driven vehicle workloads. Infotainment and application-level services are commonly deployed on platforms such as Ford SYNC~\cite{fordSYNC} and Android Automotive OS~\cite{androidAutomotiveOS}, which emphasize user interaction, application ecosystems, and connectivity. While these systems provide rich application support, they are not designed to offer the predictable execution or timing awareness required by perception and control pipelines. As a result, they are typically isolated from driving critical workloads.

Safety-critical vehicle functions are instead deployed on dedicated real-time operating systems such as QNX~\cite{qnxRTOS} and standards-based platforms like AUTOSAR Adaptive~\cite{autosar}. These systems focus on deterministic scheduling, fault isolation, and functional safety certification, making them suitable for control and actuation. However, they provide limited support for data-intensive workloads, dynamic application deployment, and flexible data access, and often treat storage and high-bandwidth sensor data as peripheral concerns.

Autonomous driving development frameworks such as ROS2~\cite{macenski2022robot}, Autoware~\cite{kato2018autoware}, and Apollo~\cite{apollo} further illustrate this fragmentation. These platforms facilitate rapid development of perception and planning algorithms, but they are built atop general-purpose operating systems and middleware that lack integrated support for real-time scheduling, coordinated data management, and system-wide safety awareness. Performance isolation, storage organization, and privacy protection are typically handled through external tools or application-level logic rather than as first-class operating system services.

Vendor-specific stacks, such as NVIDIA Drive OS~\cite{nvidiaDriveOS}, offer tight integration with specialized hardware and achieve high performance for specific deployments, but their proprietary nature limits portability, extensibility, and cross-vendor interoperability. Moreover, these platforms often prioritize compute acceleration and perception pipelines, leaving broader vehicle computing concerns such as long-term data lifecycle management, privacy-aware data sharing, and standardized programming abstractions only partially addressed.

Taken together, existing vehicle operating systems and runtimes address individual aspects of vehicle functionality in isolation. They lack a unified system foundation that simultaneously accounts for real-time driving requirements and data-centric vehicle services, motivating the need for a more integrated operating system approach tailored to the vehicle computing era.

\section{DAVOS System Architecture}
\label{section3:system}
Figure \ref{fig:system_arch} illustrates the DAVOS system architecture and how DAVOS is integrated into a modern autonomous vehicle computing architecture. As shown in Figure \ref{fig:system_arch}(b), DAVOS is deployed on the vehicle’s central computing unit, which serves as the convergence point between driving-critical Advanced Driver Assistance Systems(ADAS) / Autonomous Driving (AD) workloads and data-centric services. The central computer interfaces with multiple zonal gateways that aggregate sensors, actuators, and electronic control units distributed throughout the vehicle. Within this architecture, DAVOS provides a unified operating system layer that mediates computation, data movement, storage, and application access for both real-time driving functions and vehicle data services, while remaining agnostic to the underlying zonal or ECU layout.

Figure \ref{fig:system_arch}(a) presents the internal system architecture of DAVOS and its data flow. The architecture is organized bottom-up from hardware resources to applications, and horizontally from real-time autonomous driving workloads on the left to data-centric services on the right. This structure highlights how DAVOS supports both classes of workloads within a single operating system framework.

\textbf{Hardware \& Basic OS} At the bottom of the stack, the hardware layer consists of vehicle sensors, computing units, storage devices, and in-vehicle networks. These components generate high-rate multimodal data streams and execute compute-intensive workloads that form the foundation of both autonomous driving and vehicle data services. DAVOS is designed to operate atop existing basic operating systems, including RTOS-based variants and Linux-based systems, without replacing them.

\textbf{Runtime} Above the basic OS, the DAVOS runtime layer provides core system services that directly manage data flow and execution.
Sensor-In-Memory Communication (SIM) enables low-latency, bounded data transfer from sensors to perception and localization pipelines by maintaining sensor data in shared memory layouts that match application-native formats. This design reduces serialization overhead and timing variability in sensor-to-algorithm paths, supporting efficient data delivery for time-sensitive workloads (R1: Safety and R3: Efficiency).
Real-Time Scheduling coordinates computation and I/O execution across vehicle workloads with timing sensitivity, allowing decision-making pipelines to execute with predictable temporal behavior under varying system load (R1: Safety).

On the data-service side, Autonomous Vehicle Storage (AVS) provides hierarchical and computational storage across SSD and HDD tiers, enabling continuous data ingestion, efficient archival, and selective retrieval of vehicle data for downstream analytics (R3: Efficiency and R4: Extensibility).
Privacy-aware Confidential Computing (PaCC) integrates confidential execution and data protection mechanisms into the runtime, ensuring that sensitive vehicle data remains protected even when accessed by higher-level services or third-party applications (R2: Security).

\textbf{Management} The management layer builds on the runtime to provide system-level coordination and controlled access.
The Context-aware Risk Index (CRI) continuously evaluates environmental context to provide safety guidance for driving-critical workloads. By influencing system behavior based on assessed risk, CRI contributes to runtime safety awareness without embedding safety logic directly into applications (R1: Safety).
The Vehicle Programming Interface (VPI) exposes standardized abstractions for accessing computation, data, and system services. VPI decouples applications from hardware and runtime specifics, enabling controlled access to sensor streams, storage records, and computing resources (R4: Extensibility).

\textbf{Application} At the top, the application layer hosts both real-time ADAS or AD platforms and vehicle data services. On the left, autonomous driving frameworks execute perception, prediction, and planning pipelines using data paths and scheduling support provided by the runtime. On the right, trusted vehicle services and untrusted third-party applications perform data-driven tasks such as entertainment, diagnostics, and analytics. All application access to data and system resources is mediated through the management and runtime layers, preserving data protection and controlled extensibility.

Together, the architecture shown in Figure \ref{fig:system_arch} illustrates how DAVOS unifies real-time autonomous driving and data-centric vehicle services within a single operating system framework. By structuring system support around predictable execution, protected data handling, efficient data movement, and standardized programmability, DAVOS addresses the core requirements of Safety (R1), Security (R2), Efficiency (R3), and Extensibility (R4) in the vehicle computing era.

\section{Sensor-in-Memory Communication (SIM)}
\label{section4:sim}
\subsection{Problem Statement}
Autonomous vehicles are safety critical. As speed increases, the allowable perception to decision latency shrinks because the vehicle travels farther each millisecond. On a vehicle running the open source Autoware Universe stack, measurements show a mean perception to decision time of 521.91 milliseconds, with 369.45 milliseconds in ROS 2 publish and subscribe paths, indicating that the communication substrate can dominate end to end delay and tail behavior~\cite{liu20214c}.

Open source autonomy stacks commonly use ROS 2 with DDS for intra vehicle messaging. In this schema, publish subscribe improves modularity and reuse, but (de)serialization, redundant copies, and dynamic discovery add latency and jitter, especially with high rate cameras and LiDAR. On embedded platforms with limited CPU and memory, these DDS induced costs cap control loop frequency and compress safety margins.

Empirical studies show that message size, executor choice, and QoS settings influence mean and tail latency in ROS 2 and DDS pipelines, and that (de)serialization and redundant copying often dominate at high throughput~\cite{maruyama2016exploring, mobaiyen2022systematic, puck2021performance, kronauer2021latency, teper2022end, kouril2024performance}. Existing tuning and runtime modifications retain DDS abstractions, keeping conversion and copying on the sensor-to-application path. ROS 2 zero-copy paths reduce copies within the middleware boundary but still require converting ROS messages to algorithm-native structures such as cvMat or PCL, reintroducing overhead. DDS exposes ordering and liveliness through QoS, but per frame lifecycle management, such as sequence tags, writer heartbeats, and checksums, often remains the application’s responsibility~\cite{macenski2022robot}.

These gaps motivate SIM, Sensor-In-Memory Communication. SIM places a shared memory buffer on the sensor application path, bypassing DDS while preserving ROS 2 integration. SIM keeps sensor data in algorithm native layouts, removes (de)serialization, enforces freshness-first bounded sharing, uses constant-time lock-free writer and reader paths, and incorporates sequence identifiers, writer heartbeats, and optional checksums to provide ordering, liveness, and basic integrity. The goal is not only lower averages but tighter p95 and p99 latency in the perception to decision loop, which is directly tied to lost stopping margin at driving speed.

\subsection{Architecture Overview}

\begin{figure}[!h]
\centering
\includegraphics[width=1\columnwidth]{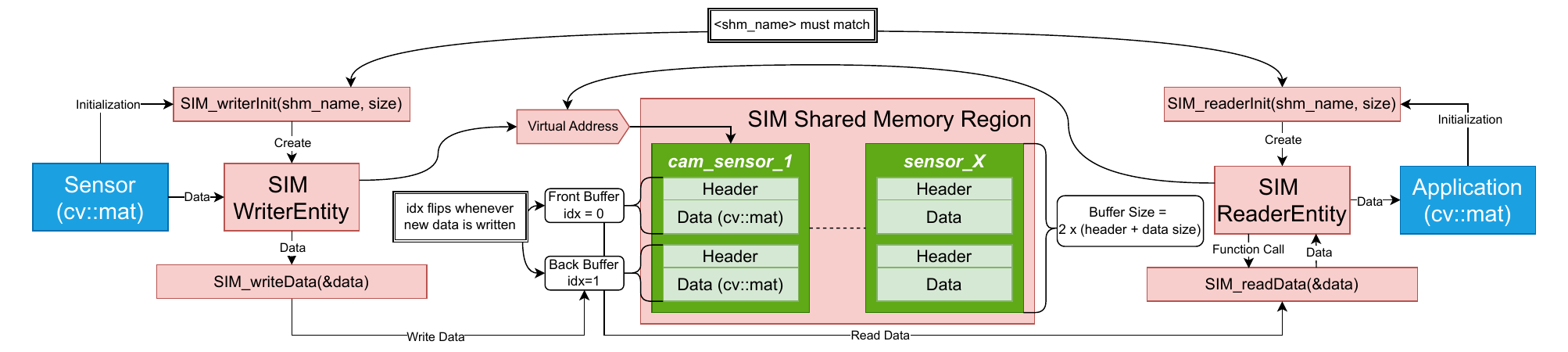} 
\caption{SIM architecture overview. SIM maps shared memory for each sensor, using unique identifiers, such as /camera\_front, with dynamically sized buffers based on sensor specifications. Each sensor has dedicated regions without disrupting others.}
\label{fig:SIM_overview}
\end{figure}

SIM provides a domain-specific shared memory communication layer that shortens the intra-host sensor-to-application path by replacing the ROS~2 and DDS message, serialization, and copy chain with a preallocated native-layout data plane. As illustrated in Figure~\ref{fig:SIM_overview}, each real-time sensor stream is assigned a uniquely named shared memory region sized from the sensor specification. Writers publish application-native data directly into this region, while readers attach by name to consume the most recent complete frame. By eliminating (de)serialization, dynamic discovery, queueing, and buffer rematerialization, SIM forms a streamlined data path that delivers predictable, low-latency transport for high-rate camera and LiDAR workloads.

\textbf{Shared memory region}. For every sensor stream, SIM creates a POSIX shared memory region once at initialization and sizes it to hold the maximum possible frame. Camera streams store width, height, channels, stride, and depth so that the payload can be viewed directly as a \texttt{cv::Mat}, while LiDAR streams allocate contiguous \texttt{PointXYZ} arrays according to a chosen upper bound on point count. All payloads are stored in the exact layout expected by downstream algorithms, removing the need for ROS message conversion or (de)serialization. Once created, the memory region is mapped and reused without any further allocation or system calls.

\textbf{Double-buffered memory layout}. Each region contains a small header and two payload buffers. The header maintains an atomic index that identifies the currently published buffer, along with per-buffer frame numbers, timestamps, and a published length. Writers always write into the inactive buffer and atomically flip the index after completing the frame, ensuring that readers never observe torn or partially written data. This bounded structure guarantees that memory use remains constant and that only complete frames are ever published.

\textbf{Lock-free publish–consume protocol}. A writer fills the back buffer completely, writes the published length, updates the frame number and timestamp, and commits the frame through an atomic index flip. Readers load the index with acquire ordering, detect whether a new frame is available by comparing frame numbers, read the published length, and copy exactly that number of points or bytes. This design ensures wait-free operation on both sides. If producers outpace consumers, intermediate frames are intentionally overwritten, implementing SIM’s freshness-first policy that favors timely decision-making in autonomous driving.

\textbf{Capacity provisioning and variable-size handling}. Although each region is sized for the sensor’s maximum frame, SIM uses the published length field to describe the actual amount of data written. Writers set this value before publishing, and readers observe it under acquire semantics. This mechanism eliminates per-frame size races and maintains constant-time publication regardless of frame variability, such as changes in LiDAR point counts.

\textbf{Correctness and timing model}. Correctness follows from a single writer per stream, bounded double buffers, write-complete-before-publish ordering, and atomic index updates. These invariants ensure readers never encounter inconsistent or out-of-bounds data. Steady-state handoff time is dominated by memory-copy bandwidth and a small number of atomic operations. Because SIM avoids discovery, dynamic allocation, and (de)serialization, its transport latency exhibits both lower mean and substantially tighter p95 and p99 tails.

\textbf{OS-aware placement}. Region headers and payloads are cache-line aligned, and both writers and readers are pinned to the same NUMA node to improve locality. Writers map the region read–write, readers map it read-only, and high-rate streams may lock their memory pages to avoid paging delays. These considerations reduce jitter and contribute to SIM’s predictable timing.

\textbf{Compatibility and deployment model}. SIM integrates into existing ROS~2 nodes with minimal changes, typically around four lines per publisher–subscriber pair. Bridge processes can expose SIM streams to remote DDS consumers or mirror DDS inputs into SIM, enabling incremental adoption while preserving inter-ECU communication via ROS~2 and DDS. SIM intentionally uses double buffering rather than ring buffers because queues increase delay and tail variance and optimize for completeness rather than freshness. SIM remains strictly an intra-host transport with a single writer per stream, aligning with its scope and design assumptions.

\section{Real Time Scheduling}
\label{section5:schedule}
\subsection{Problem Statement}

Ensuring predictable timing behavior for autonomous driving remains one of the most difficult challenges in vehicle computing. Modern autonomy stacks contain deeply pipelined perception, prediction, and planning workloads that interact across heterogeneous sensors, asynchronous data flows, and multiple hardware domains. Their timing behavior cannot be accurately captured by traditional real-time abstractions, which assume discrete triggering, uniform task dependencies, and simple deadline semantics~\cite{casini2019response, choi2021picas, durr2019end}. Instead, these pipelines generate temporally continuous results, maintain histories for prediction, and output trajectories whose validity degrades over time rather than at a single deadline~\cite{wu2024timeliness}.

At the same time, contemporary vehicle architectures divide computation across high-performance central compute units and smaller real-time domain controllers~\cite{bandur2021making}. This produces two execution paths: a hard real-time bypass responsible for collision-avoidance safety~\cite{yu2020building}, and a soft real-time main autonomy stack responsible for high-level decision-making. The main stack, therefore, does not need to meet deadlines with strict determinism, but it must continuously maintain sufficient quality in perception, prediction, and planning outputs to ensure stable and safe driving behavior~\cite{wu2024timeliness}.

For these reasons, DAVOS adopts a real-time philosophy centered not on rigid deadline satisfaction, but on maintaining the \emph{confidence} of the final autonomy outputs. The goal of scheduling is to ensure that the overall perception–planning pipeline operates with high informational integrity even under load spikes, data delays, or resource contention, while coexisting with other DAVOS components such as AVS, CRI, PaCC, and third-party applications. This shift from deadline-driven correctness to confidence-centric correctness motivates a novel design of the DAVOS real-time scheduler.

\subsection{Architecture Overview}

\begin{figure}[!h]
\centering
\includegraphics[width=.9\columnwidth]{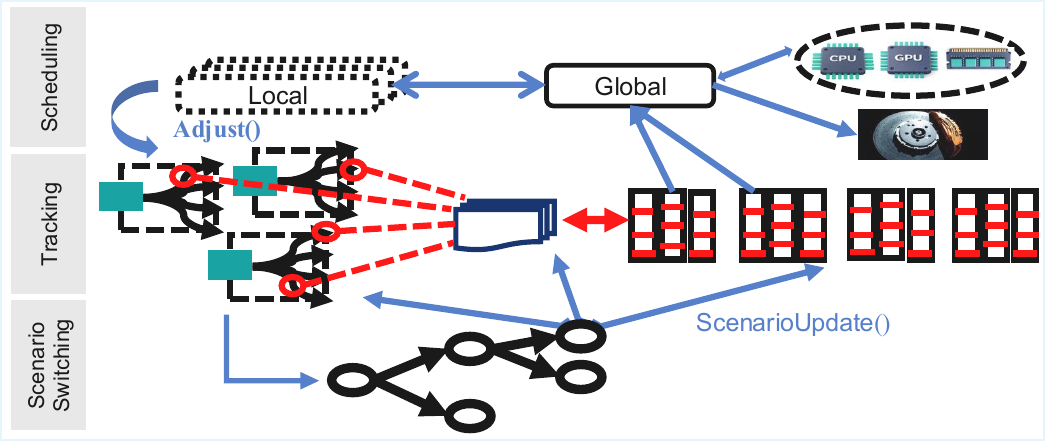} 
\caption{Real-time scheduling architecture overview. It contains three parts: scenario switching, tracking, and scheduling (local and global).}
\label{fig:Scheduling_overview}
\end{figure}

The DAVOS real-time scheduling subsystem implements a confidence-driven runtime architecture composed of three cooperating components shows in Figure~\ref{fig:Scheduling_overview}: the tracking component, the scheduling component, and the scenario-switching component. These components operate over the unified subsystem structure used throughout DAVOS, where each subsystem is modeled as a single-source, multi-output execution graph with active dependencies and well-defined temporal semantics.

The tracking component continuously monitors confidence across critical paths, converting data staleness, fusion misalignment, and execution delays into quantitative indicators. The scheduling component coordinates computation across both local and global scopes: each subsystem owns a local scheduler responsible for shaping its internal execution, while a global scheduler resolves overload, reallocates cores, and maintains system-wide balance. The scenario-switching component adapts the runtime according to driving context, ensuring that confidence computation, monitoring thresholds, and prioritization strategies match the requirements of the selected scenario.

\textbf{Scenario Switching.}
Different driving scenarios impose different timeliness and accuracy requirements. For example, lane-change maneuvers emphasize lateral-object tracking, highway cruising emphasizes long-range prediction, and parking emphasizes fine-grained perception. The scenario-switching component updates runtime parameters whenever the planning module selects a new scenario, including confidence computation weights, monitoring thresholds, path priority orders, and downgrade permissions. It may also adjust subsystem rates to better match the timing demands of the new context. Scenario switching provides anticipatory adaptation, reducing the likelihood of timing problems and improving overall determinism.

\textbf{Tracking.}
The tracking component computes confidence at strategic monitoring points across the autonomy pipeline. Confidence reflects both the quality of input samples and the temporal alignment between fused sensor data. Because instantaneous confidence may fluctuate due to normal execution jitter, the system maintains long-horizon indicators such as smoothed averages, lower quantiles, and (m,k)-style satisfaction metrics. When these indicators degrade past predefined thresholds, the system generates regulation requests; when they recover, restoration requests are issued. This mechanism unifies timing failures, data irregularities, and fusion inconsistencies into a common confidence-based signal.

\textbf{Scheduling.}
The scheduling component operates in a two-level hierarchy. Each subsystem includes a local scheduler that uses active dependencies, instance ordering, and slack-based priority shaping to optimize execution. When regulation is required, the subsystem progressively reduces the effective deadlines of affected paths or activates application-level degradation strategies such as reduced-fidelity models or controlled skipping. If local adjustments are insufficient, responsibility escalates to the global scheduler. The global scheduler manages isolation and resource distribution across subsystems, reallocating CPU cores or adjusting execution budgets to ensure that critical paths recover. This division of responsibilities balances subsystem autonomy with system-wide stability.





\section{ Context-aware Risk Index (CRI)}
\label{section6:cri}
\subsection{Problem Statement}

Ensuring safety in autonomous driving requires precise, real-time estimation of environmental risk and adaptive behavioral control. However, existing risk estimation approaches suffer from several fundamental limitations. Many methods produce coarse, global scene-level metrics with limited interpretability~\cite{yu2021dynamic}, while others introduce risk indicators without concrete integration into autonomous systems~\cite{liu2019towards} or are narrowly designed for specific driving scenarios such as overspeed detection or lane-changing~\cite{kutela2025influence, naveiro2024adversarial}. These constraints make it difficult for an AV stack to understand the direction, type, and severity of risk in a way that can drive real-time control decisions.

Moreover, prior work typically treats risk as a scalar quantity, ignoring that driving risk is inherently directional and object-specific. Existing complexity or scenario-based frameworks rarely incorporate kinematic relationships such as relative orientation, directional closing speed, or time-to-collision. Other safety-envelope approaches, such as probabilistic risk bounds~\cite{bernhard2022risk} or RSS conservative braking rules~\cite{shalev2017formal}, provide useful constraints but lack fine-grained, context-sensitive modeling of dynamic interactions between the ego vehicle and surrounding objects.

These gaps motivate the Context-aware Risk Index (CRI), a lightweight, modular risk-assessment runtime designed for integration directly into an autonomous driving control loop. CRI quantifies directional risks by analyzing object kinematics, spatial relations, and safety envelopes, producing interpretable risk scores that reflect localized threats. Rather than generating a single global value, CRI yields direction-aware risk distributions, enabling autonomous systems to dynamically adjust throttle, brake, and steering to safer modes. CRI thereby provides a practical runtime mechanism for real-time, risk-aware driving behavior adaptation, addressing the lack of interpretable, directional, and control-integrated risk estimation in prior literature.

\subsection{Architecture Overview}

\begin{figure}[!h]
\centering
\includegraphics[width=1\columnwidth]{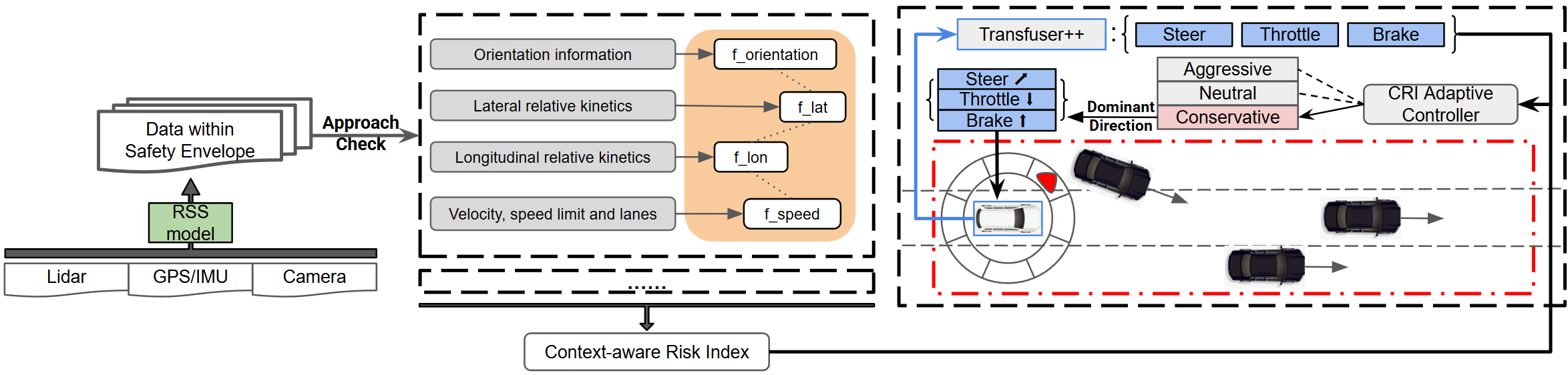} 
\caption{System overview of CRI integration.CRI computes directional risk from RSS-filtered objects and adjusts Transfuser++ control via aggregated value and dominant direction.}
\label{fig:CRI_overview}
\end{figure}

The Context-aware Risk Index (CRI) operates as a modular runtime integrated into the DAVOS Management Layer, where it continuously evaluates environmental and vehicle-state factors to guide safe control decisions. As illustrated in the Figure~\ref{fig:CRI_overview}, the architecture transforms raw perceptual data into directional, kinematic-aware risk signals and uses them to modulate control in real time. The full pipeline consists of three components: per-object risk computation, direction-aware aggregation, and adaptive control integration.

\textbf{CRI Calculation Pipeline.} At each decision cycle, CRI begins by constructing a dynamic safety envelope using the Responsibility-Sensitive Safety (RSS) model. The envelope’s longitudinal safe distance is determined by RSS braking equations, while lateral boundaries adjust with lane counts. Only objects inside this envelope are considered for risk computation.

For each detected object, CRI computes three independent risk components based entirely on the CRI paper’s formulation:
(1) Orientation risk, derived from the relative orientation angle between the ego and the object following Montewka et al.’s angular collision model.
(2) Directional velocity risk, encapsulating longitudinal and lateral Time-to-Collision (TTC) using an approach-check mechanism: objects whose relative velocity indicates receding movement contribute zero risk, while approaching objects produce decaying TTC-based risk factors exponentially.
(3) Speed-limit risk, modeled using a logistic function incorporating ego speed, road speed limits, and lane count following Kutela et al.

These factors are fused via a probabilistic-max hybrid strategy, ensuring both cumulative interaction sensitivity and prioritization of extreme threats. The resulting per-object risk scores are bounded, intuitive, and interpretable. And the corresponding CRI's equation is :

\begin{align}
\text{CRI} = \left[ \alpha \cdot f_{\text{spatial}} + (1-\alpha) \cdot \max_i f_i \right] \cdot \frac{e^{f_{speed} - \text{speed\_ref}}}{e^{\text{speed\_ref}}}
\end{align}

where \( f_{\text{speed}} \in [0,1] \) reflects the ego vehicle's speed-related risk, and \( \text{speed\_ref} \) is a calibrated neutrality point. And \( f_{\text{spatial}}\) is spatial risk.

\textbf{Direction-aware CRI Aggregation.} To preserve spatial relationships, CRI partitions the space around the ego vehicle into eight equally spaced sectors. Each object’s CRI value is assigned to its corresponding sector, where the maximum risk within that sector is retained. A hybrid vector-max fusion then generates a scene-level risk magnitude by combining directional maxima with smooth spatial integration. The algorithm simultaneously identifies the dominant risk direction, which will later influence control adjustments.

The aggregated risk vector magnitude \( R_{\text{vector}} \) is:
\begin{align}
R_{\text{vector}} = \sqrt{\left(\sum_{d} R_d \cos \theta_d\right)^2 + \left(\sum_{d} R_d \sin \theta_d\right)^2}
\end{align}
where \( \theta_d \) is the central angle of sector \( d \). In parallel, the maximum directional risk is identified as \( R_{\text{max}} = \max_d R_d \).

The final aggregated CRI score is obtained through a weighted sum:
\begin{align}
\text{CRI}_{\text{final}} = \beta R_{\text{vector}} + (1-\beta) R_{\text{max}}   
\end{align}
where \( \beta = 0.7 \) balances between smooth spatial integration and peak risk sensitivity.

\textbf{Adaptive Control Integration.} The final CRI signal consists of the aggregated magnitude and the dominant directional angle. These are fed directly into the autonomous driving control layer (e.g., Transfuser++ in the CRI evaluation) to modulate driving style. The controller transitions between predefined aggressive, neutral, or conservative modes depending on CRI magnitude and risk direction. This integration allows CRI to enforce proactive deceleration, smoother motion, and early hazard response. Importantly, CRI’s design is modular and lightweight, adding only 3.6 ms per decision cycle with no need for model retraining.

The corresponding control logic is summarized as follows:

\begin{algorithm}[H]
\caption{CRI-Based Adaptive Control Policy}
\begin{algorithmic}[1]
\State \textbf{Input:} Ego state $s_{ego}$, detected objects $\mathcal{O} = \{o_i\}$
\State \textbf{Initialize:} Risk calculator $\mathcal{C}$, adaptive controller $\mathcal{A}$
\While{navigation is ongoing}
    \State Update object list $\mathcal{O}$
    \For{each $o_i \in \mathcal{O}$}
        \State Compute directional risk $r_i = \mathcal{C}(o_i)$
    \EndFor
    \State Aggregate risks into distribution vector $\mathbf{r}$
    \State Identify dominant risk direction $\theta^*$ and risk magnitude $r^*$
    \State Select driving mode $m \leftarrow \mathcal{M}(r^*, \theta^*)$
    \State Compute adapted control $u_{control} = \mathcal{A}(s_{ego}, m)$
    \State \textbf{Return} $u_{control}$
\EndWhile
\end{algorithmic}
\end{algorithm}

Through this architecture, CRI bridges perception and control with fine-grained, direction-aware risk understanding, yielding significantly safer and more stable behavior across adverse and failure-prone scenarios, exactly as demonstrated in the original CRI evaluation.

\section{Autonomous Vehicle Storage System (AVS)}
\label{section7:avs}
\subsection{Problem Statement}
As vehicles evolve into intelligent computing platforms, onboard data has become a critical resource for autonomy, system intelligence, and post-drive analytics. Each vehicle continuously produces massive heterogeneous data streams from cameras, LiDAR, radar, GNSS, IMU, and CAN, often reaching terabytes per day~\cite{wang2024quantitative}. However, most of this data is ephemeral: it is consumed for immediate control and then discarded, leaving no persistent, queryable foundation for future use.

Emerging applications increasingly depend on historical sensor data. For example, safety and forensic analysis require precise temporal replay around events, fleet analytics rely on aggregated long-term logs, and machine learning pipelines draw on diverse sensor histories to mine rare or corner cases. These workloads demand a storage substrate that can support the full lifecycle of data, from continuous high-rate ingestion to flexible query and long-term retention within limited embedded resources.


Existing in-vehicle storage stacks fail to meet these requirements. Conventional logging tools, such as ROS 2 bag and MCAP, are optimized for sequential recording and offline replay rather than for continuous, queryable operation. Their append-only formats lead to unpredictable latency, limited concurrency, and inefficient retrieval~\cite{xu2024rosfs}. Prior research systems such as HydraSpace focus primarily on compression and lack validated indexing and archival pipelines~\cite{wang2020hydraspace}. General edge and IoT storage designs assume networked clusters and replication, assumptions that do not hold within a single vehicle with bounded capacity and power~\cite{cai2016iot, meruje2024databases}.

This gap motivates AVS (Autonomous Vehicle Storage), a computational and hierarchical storage architecture that integrates modality-aware reduction and compression, hot and cold tiering, and lightweight metadata indexing into a unified design. AVS treats onboard storage as a first-class system service, enabling predictable ingest, efficient retrieval, and long-term retention to support the data foundation of vehicle computing.

\subsection{Architecture Overview}
\begin{figure}[!h]
\centering
\includegraphics[width=1\columnwidth]{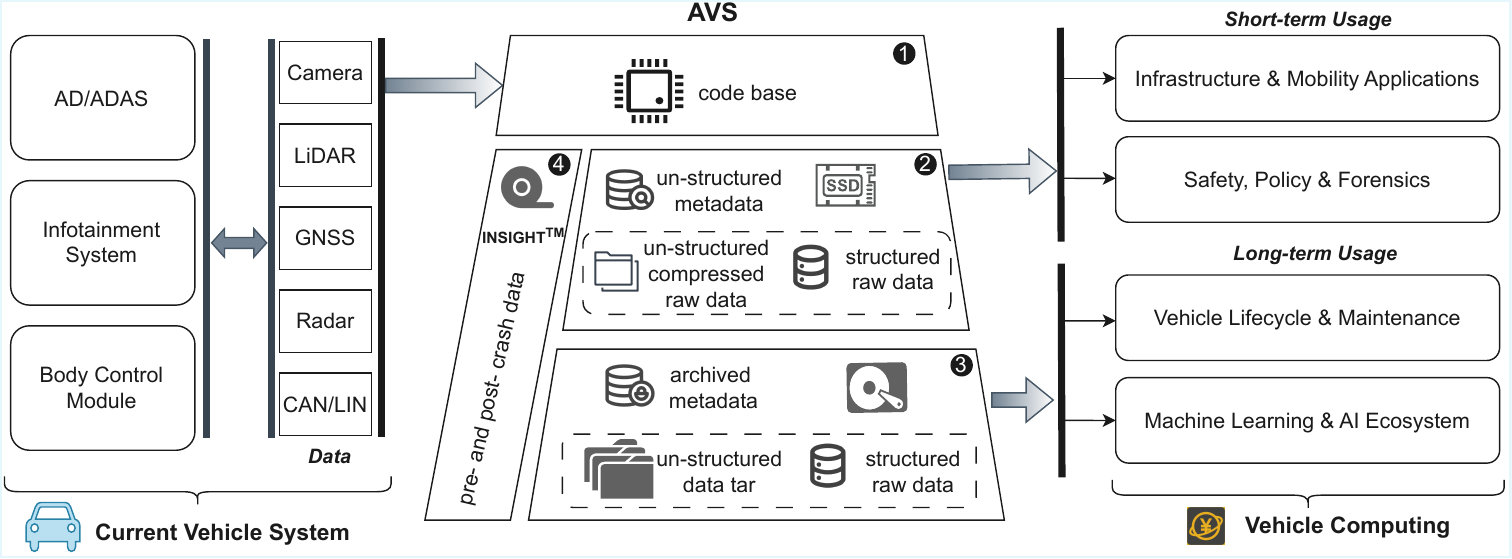} 
\caption{The autonomous vehicle storage system (AVS) architecture. AVS is separate from the real-time ADAS/AD system, with all computation running on separate computing units. It contains four parts: code base computing, SSD for hot storage, HDD for cold storage, and a tap for incident data collection.}
\label{fig:AVS_overview}
\end{figure}

Figure~\ref{fig:AVS_overview} illustrates the overall design of the Autonomous Vehicle Storage (AVS) system. AVS redefines onboard data management through a computational and hierarchical architecture that unifies data reduction, compression, indexing, and tiered storage within a single runtime pipeline. The design is guided by four key principles:

(A) Heterogeneity-aware ingestion. Vehicle workloads combine high-bandwidth sensors such as LiDAR and cameras with low-bandwidth telemetry from GNSS, IMU, and CAN. The storage path must accommodate different data rates and formats concurrently without frame loss or cross-stream interference. AVS decouples ingestion paths from format-specific bottlenecks and provides predictable latency for each sensor stream.

(B) Computational and hierarchical layout. Onboard storage is inherently limited in capacity. AVS integrates lightweight, streaming-friendly reduction and compression modules to minimize data footprint while maintaining downstream utility. It further organizes storage into a two-tier hierarchy: a solid-state drive (SSD) tier for short-term access and a hard-disk drive (HDD) tier for archival retention. This structure balances real-time write throughput with long-term durability.

(C) Query-friendly organization. Beyond raw replay, many applications require selective retrieval of sensor data based on time or modality. AVS introduces lightweight metadata indexing that records sensor identifiers, timestamps, and file paths, enabling range queries and fine-grained lookup. This design allows fast retrieval of both structured (e.g., GNSS) and unstructured (e.g., LiDAR, image) data without scanning entire files.

(D) Lightweight and resource-aware operation. Vehicle compute platforms operate under strict power and thermal budgets. Every AVS component—from compression and indexing to data movement—is designed as a streaming process with bounded memory and CPU usage. The system achieves predictable performance on embedded devices while running concurrently with the autonomy stack.

The AVS architecture consists of three coordinated layers:

\textbf{Real-time ingestion layer}. This layer hosts the AVS core services, including data reduction, compression, archival movement, and retrieval. Incoming sensor data is processed and written to the SSD tier in real time. When the vehicle is idle, such as overnight, the archival service migrates older data from SSD to HDD. The retrieval interface exposes time- and modality-based access to recent histories with bounded latency.

\textbf{Hot tier (SSD)}. The SSD stores the most recent and frequently accessed data. Unstructured data such as images and LiDAR scans are first reduced and compressed before being stored in day-organized directories, with metadata entries recorded in an embedded SQLite database for fast lookup. Structured data such as GPS and CAN messages are written directly into per-day databases. This tier ensures high ingest throughput and immediate queryability.

\textbf{Cold archive tier (HDD)}. The HDD provides long-term storage for historical data. Its directory layout mirrors that of the SSD but is organized hierarchically by year and month (YYYY/MM) to prevent directory bloat. Before transfer, unstructured files are packed into tar archives to reduce fragmentation and improve sequential access efficiency. Archival metadata maintains consistency between tiers, logging each migration event’s time, file count, and data range.

\textbf{INSIGHT\texttrademark: Intelligent System for Incident Gathering, Handling, and Tracing}.
INSIGHT functions as the crash-aware forensic extension of AVS, responsible for safeguarding mission-critical evidence during and after abnormal driving events. Built upon the AVS ingestion services, it continuously monitors vehicle dynamics and sensor feedback to detect potential crash events in real time. Upon a confirmed incident, INSIGHT locks a pre- and post-crash time window from the SSD buffer and performs a write-once archival into the subsystem, ensuring tamper-proof preservation of all related data. Beyond storage, INSIGHT supports synchronized multi-sensor alignment and scene reconstruction, enabling precise post-event analysis for diagnostics, liability investigation, and safety model improvement.

Together, these layers form an end-to-end storage pipeline capable of real-time ingestion, efficient querying, and space-efficient archival within a fixed onboard footprint. By coupling computation with hierarchical organization, AVS transforms the storage subsystem from a passive logger into an active data service that sustains both autonomous driving and long-term vehicle data analytics.

\section{Vehicle Programming Interface (VPI\texttrademark)}
\label{section8:vpi}
\subsection{Problem Statement}
Developing applications for vehicle computing remains difficult due to the interdisciplinary complexity of autonomous driving, which spans computer vision, machine learning, sensor fusion, control, and software engineering, creating a high technical barrier for developers~\cite{padmaja2023exploration,liu2020computing}. The problem is compounded by heterogeneity across manufacturers: vehicles differ widely in sensor types, data formats, and system architectures, making applications hard to generalize across platforms~\footnote{Goncharov I. Autonomous vehicle companies and their ML, 2013. \url{https://wandb.ai/ivangoncharov/AVs-report/reports/Autonomous-Vehicle-Companies-And-Their-ML-VmlldzoyNTg1Mjc1}, Dec. 2025}. 

Existing autonomous driving stacks further increase development complexity. Systems such as Autoware contain hundreds of ROS nodes, illustrating the difficulty of integrating large numbers of components and their dependencies~\cite{macenski2022robot}. At the same time, security and safety concerns push companies toward closed, vehicle-specific software, which limits collaboration and prevents the growth of an open vehicle computing ecosystem~\cite{pham2021survey,sun2021survey}.

Although industry solutions provide partial support, substantial gaps remain. AUTOSAR focuses mainly on communication and ECU-level control rather than high-level computational or data-driven tasks~\cite{furst2016autosar}. Other ecosystems—such as Autoware~\footnote{\url{https://autoware.org/}, Dec. 2025}, Apollo~\footnote{\url{https://github.com/ApolloAuto/apollo}, Dec. 2025}, SOAFEE~\footnote{SOAFEE Architecture. 2023. \url{https://architecture.docs.soafee.io/en/latest/contents/introduction.html}, Dec. 2025}, NVIDIA DRIVE~\footnote{NVIDIA. NVIDIA DRIVE end-to-end solutions for autonomous vehicles. 2023. \url{https://developer.nvidia.com/drive}, Dec. 2025}, and IVY~\footnote{\url{https://www.blackberry.com/us/en/products/automotive/blackberry-ivy#features}, Dec. 2025}. They offer APIs within their own frameworks but do not provide a unified, cross-platform programming model for vehicle computing.

To address these challenges, VPI proposes the first comprehensive, standardized programming interface suite for vehicle computing. By abstracting hardware, data, computation, service, and management functions into modular APIs, VPI reduces development effort, improves portability across vehicle platforms, and enables an open, interoperable software ecosystem for connected and autonomous vehicles.


\subsection{Architecture Overview}

\begin{figure}[!ht]
\centering
\includegraphics[width=1\columnwidth]{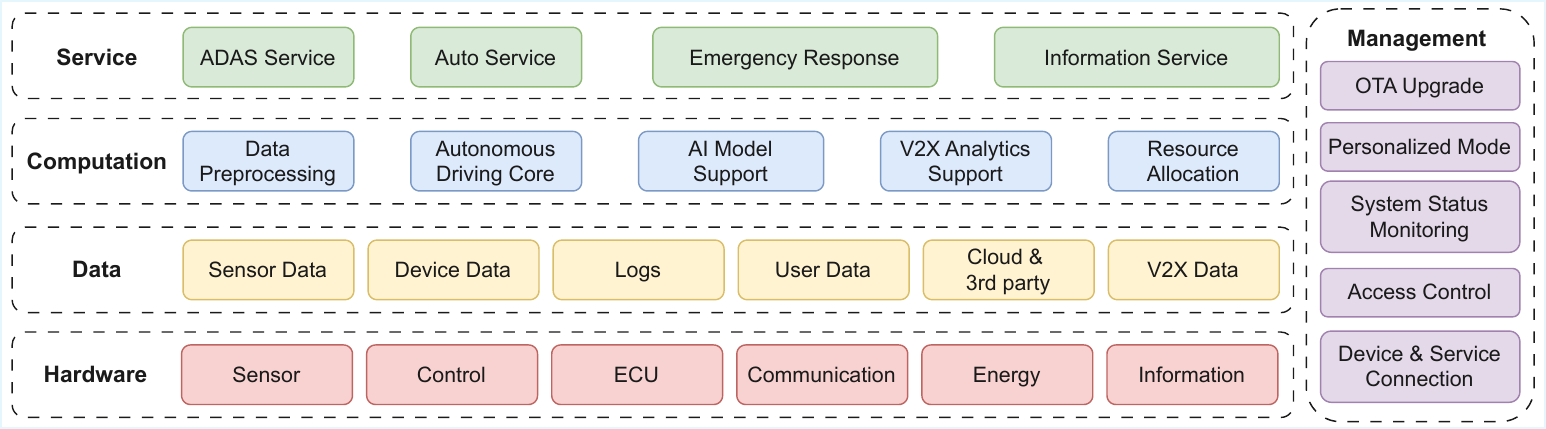} 
\caption{Structure of VPIs. VPIs are composed of five main categories: Hardware, Data, Computation, Service, and Management.}
\label{fig:VPI_overview}
\end{figure}

Figure~\ref{fig:VPI_overview} presents the overall architecture of the Vehicle Programming Interface (VPI). The architecture is built on the principle of layered abstraction, defining five categories of standardized interfaces that collectively span the entire vehicle software stack: Hardware, Data, Computation, Service, and Management. This design provides a structured pathway for developers to access vehicle resources, process data, invoke computation, and manage applications with consistent semantics across platforms.

\textbf{Hardware VPIs.}
The Hardware VPI abstracts the physical components of the vehicle, providing a unified interface for sensors, actuators, controllers, communication devices, energy systems, and infotainment modules. It hides hardware differences across manufacturers and enables direct access to sensor configuration, actuator control, ECU management, and power operations. Sub-components include Sensor VPI for enumeration, calibration, and configuration; Actuator VPI for motion control; ECU VPI for subsystem coordination; Communication VPI for cellular, Wi-Fi, and V2X connectivity; Energy VPI for battery charging and power output; and Infotainment VPI for display and audio configuration. Together, they enable standardized, hardware-agnostic access to the vehicle’s embedded devices.

\textbf{Data VPIs.}
The Data VPI unifies the acquisition, storage, and synchronization of both real-time and historical data generated by the vehicle and its environment. Built primarily upon the AVS architecture, it provides standardized APIs for accessing sensor data, device status, user information, operational logs, infotainment content, and V2X or cloud-based sources. Through these interfaces, applications can efficiently query, analyze, and fuse multimodal information from cameras, LiDAR, GNSS, user profiles, and external services. This abstraction enables upper-layer systems to operate over a consistent data model, supporting use cases from perception and diagnostics to personalized in-cabin intelligence.

Within the data service layer of vehicle computing, the Data VPI also enforces authentication and access control. For untrusted or third-party applications, data retrieved from AVS is routed through the PaCC layer for sanitization and privacy preservation before release, ensuring secure, verifiable, and policy-compliant data access across the DAVOS ecosystem.

\textbf{Computation VPIs.}
At the computational layer, VPI exposes the vehicle’s processing capabilities as a programmable resource pool. It supports data preprocessing, AI inference, perception fusion, and task offloading across the vehicle, roadside infrastructure, or cloud nodes. Core sub-components include Data Preprocessing VPI for filtering and formatting, Autonomous Driving Core VPI for sensor fusion and path planning, AI Model VPI for inference execution, V2X Analytics VPI for collaborative computing, and Resource Allocation VPI for dynamic CPU, GPU, and accelerator scheduling. These interfaces provide a standardized mechanism for computational orchestration, ensuring that workloads such as obstacle detection, trajectory planning, and behavioral prediction are executed efficiently and adaptively.

\textbf{Service VPIs.}
Service interfaces encapsulate the vehicle’s functional modules into callable services that support rapid development of high-level applications without exposing the complexity of underlying systems. This category includes Advanced Driver Assistance Service VPI for safety-critical features such as lane keeping and adaptive cruise control, Autonomous Driving Service VPI for mode management and V2X-enhanced automation, Emergency Response VPI for incident handling and alerting, and Infotainment Service VPI for media and interaction control. These services collectively simplify the creation of application logic for autonomous driving, passenger assistance, and user experience enhancement.

\textbf{Management VPIs.}
The Management VPI provides system-level control and supervision for vehicle security, configuration, and lifecycle management. It integrates Device and Service Connection VPI for pairing and cloud access, Access Control VPI for authentication and authorization, System Monitoring VPI for real-time health and resource tracking, Personalization VPI for user-specific modes and preferences, and OTA Update VPI for secure software maintenance. This layer ensures reliability and safety through continuous monitoring, controlled access, and adaptive configuration of the entire system.

Together, these five categories form a comprehensive interface suite that transforms vehicles into open and programmable computing platforms. By standardizing access to hardware, data, computation, services, and management functions, the VPI framework decouples application logic from platform-specific implementations, paving the way for an extensible and interoperable vehicle computing ecosystem under the DAVOS architecture.

\section{Privacy-aware Confidential Computing}
\label{section9:privacy}
\subsection{Problem Statement}
Modern vehicle computing platforms support not only trusted real-time ADAS and vehicle control, but also a growing ecosystem of untrusted third-party applications that request access to historical data. Typical examples include usage-based insurance scoring, road surface and work zone mapping, fleet health analytics, incident reconstruction, personalized infotainment and advertising, and public safety requests such as locating a missing person~\cite{consumerreports2025ubi, geotab2025ubi, geotab2025fleet, nyulawreview2023autodata, gizmodo2021onstarice}. The data these applications request is rich and longitudinal. It includes pedestrian and passenger faces, vehicle license plates, driver voice and conversation audio, in-cabin gestures, device identifiers, VIN, high-resolution imagery and point clouds, GNSS traces with home and workplace inference, and even health or fatigue proxies from biosignals and camera cues~\cite{dxc2023anonymization, gallio2023avprivacy, forrester2024carlistening, pmc2024drivermonitoring, liu2022fatigue, pmc2025biosensing, insidegnss2018privacy}. Much of this data also contains bystanders who never consented to collection~\cite{eff2023avprivacy}.

The core tension is clear. Third-party applications often need information derived from the data, but rarely need the raw data itself. Without strong technical controls, granting access to raw records enables overcollection, secondary use, data linkage, and long-term retention beyond the intended purpose~\cite{pese2023pricar}. A privacy-aware confidential computing substrate is needed to enforce least disclosure, to provide verifiable processing of sensitive records, and to return only the minimal results required by a declared purpose while meeting the latency and reliability constraints of vehicle computing.

Current practice in vehicle and edge systems relies on encryption at rest and in transit, and account-scoped tokens and access control lists. Although real-time blurring solutions exist for specific objects such as faces or license plates, these approaches remain task-specific and fail to generalize into a unified framework for privacy-preserving data sharing~\cite{spyne2025, celantur2025}. Transport security in ROS 2 and DDS~\cite{mayoral2022sros2}, and security profiles in automotive standards~\cite{ISO21434}, protect communication but do not constrain how applications use plaintext once access is granted. 
Techniques such as differential privacy~\cite{dwork2006differential} address aggregate statistics but not individual media release or event retrieval. Fully homomorphic encryption~\cite{gentry2009fully} remains impractical for real-time multimodal analytics. As a result, once data is decrypted in application memory, there is no verifiable guarantee that only the necessary features are computed, no assurance that code runs in an attested environment bound to a stated purpose, and no comprehensive audit of what was derived, disclosed, and retained.

\subsection{Threat Model}





PaCC assumes a realistic in-vehicle threat model where the vehicle's operating system, middleware, and third-party software may be partially compromised. Within this environment, we assume the following components remain trustworthy:

\begin{itemize}
    \item The hardware trusted execution environment (TEE) that hosts the confidential execution containers.
    \item The PaCC runtime, including the trusted key manager and data anonymization pipeline
    \item The initial boot process and secure firmware.
    \item Cryptographic primitives that follow modern standards.
\end{itemize}

PaCC does not aim to defend against physical destruction of hardware, side-channel attacks on the cryptographic engine, or compromises of the vehicle manufacturer's cloud infrastructure.

\textbf{Attacker Ability.}
The adversary may gain elevated privileges on the host system and is capable of inspecting or modifying stored data on the compromised platform. The attacker can inject unauthorized applications with malicious intent and observe system memory, file system content, interprocess communication (IPC), and external storage devices connected to the vehicle.

However, the attacker cannot break standard encryption algorithms, cannot extract confidential state from execution containers protected by the TEE, and cannot access private cryptographic keys that never leave the trusted domain.

\textbf{Attacker Goal.}
The attacker's primary objectives are to:

\begin{enumerate}
    \item Compromises user privacy by accessing sensitive personal data generated or stored within the vehicle
    \item Bypass security controls to gain unauthorized access to protected information
    \item Manipulate or exfiltrate data for malicious purposes such as surveillance, identity theft, or unauthorized profiling
\end{enumerate}

Within this threat model, PaCC protects user privacy and restricts untrusted applications by ensuring that all stored data is encrypted, all application computation occurs in isolated confidential execution containers, and all access requests pass through strict policy enforcement and privacy-aware data reduction mechanisms.

\subsection{Architecture Overview}

\begin{figure}[!h]
\centering
\includegraphics[width=1\columnwidth]{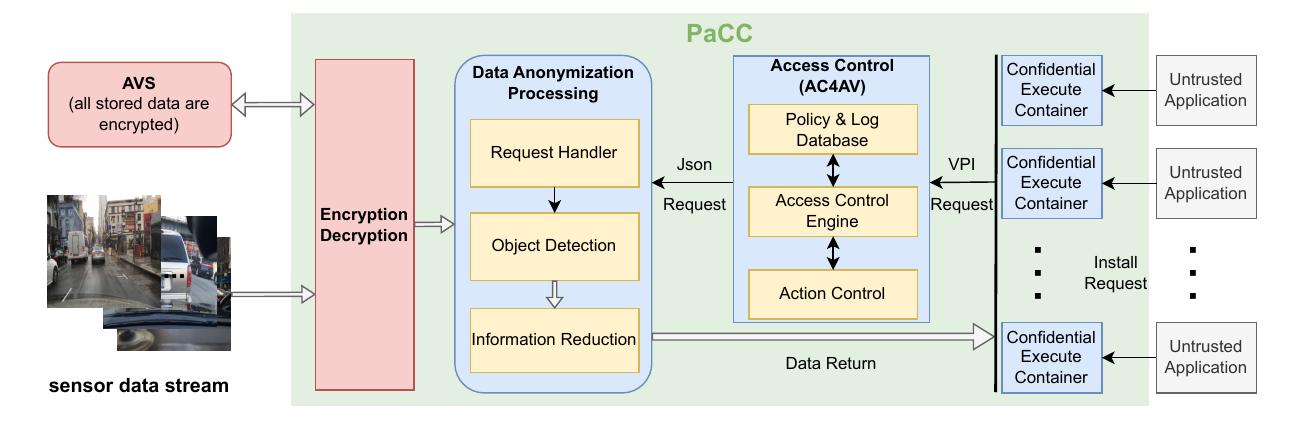} 
\caption{PaCC system architecture. Green blocks represent core PaCC functions, red blocks represent algorithms run on AVS hardware, and blue blocks represent processing that happens on PaCC Hardware.}
\label{fig:PaCC_overview}
\end{figure}

Figure \ref{fig:PaCC_overview} presents the system architecture of PaCC. The design consists of four core modules that work together to provide privacy-preserving and confidential data access on the vehicle computing platform: data encryption and decryption, data anonymization, access control, and confidential execution.

On the data ingestion path, PaCC enforces encryption on all incoming sensor streams before they are stored in AVS. A vehicle-owned private key is used to encrypt the data at the edge, ensuring that every record stored in AVS remains protected even if the operating system or storage stack is compromised. The AVS computation path (red blocks in the figure) handles ingestion and encrypted storage only; PaCC does not allow raw data to persist in any unprotected form.

On the application side, every untrusted third-party application must first be approved by the user and scanned by the system security services. Once approved, the system provisions a dedicated confidential execution container for the application. All computation of the application must take place inside this trusted execution environment, where memory, execution state, and internal storage remain isolated from the host.

During runtime, applications access data through the Vehicle Platform Interface (VPI). When an application submits a VPI request, the request is routed to the PaCC Access Control subsystem. This subsystem contains three components: the Policy and Log Database, the Access Control Engine, and the Action Control module.

The Policy and Log Database stores application registration information, audit logs, user policies, and past access decisions.
The Access Control Engine evaluates each request by checking application identity, requested data type, temporal and spatial scope, and the user defined policy rules.
The Action Control module enforces the final decision by either granting the request or returning a denial.

After a request is approved, the Access Control subsystem generates a structured JSON instruction for the Data Anonymization Pipeline. This pipeline first retrieves and decrypts the corresponding data from AVS inside the PaCC trusted domain. It then performs privacy preserving transformations that match the request, including object detection, face and license removal, spatial cropping, attribute filtering, and content reduction. Only the anonymized content is returned to the application’s confidential execution container.

Through this design, PaCC ensures that even if the host system is compromised, all data stored in AVS remains encrypted, and all data processed by each application remains isolated and privacy filtered. PaCC protects the privacy of vehicle users while allowing approved third-party services to access only the minimum necessary information. This architecture provides strong security at the storage level, execution level, and application level.

\section{Summary}
\label{section10:conclude}
Autonomous vehicles increasingly operate as both real-time cyber–physical systems and data-centric computing platforms. However, existing vehicle operating systems are typically developed around individual functions, leading to fragmented support for driving-critical workloads and vehicle data services. This separation introduces inefficiencies in data movement, execution coordination, and resource usage, and complicates the interaction between autonomy pipelines and data-driven applications. DAVOS is designed as a unified vehicle operating system architecture that brings these roles together within a common system foundation, aligning real-time autonomy with the broader vehicle computing ecosystem.

On the real-time side, DAVOS focuses on predictable execution and timely data delivery for autonomous driving workloads. Sensor-In-Memory Communication (SIM) provides a bounded and low-latency data path between sensors and perception pipelines, while real-time scheduling supports consistent execution of perception, prediction, and planning under varying system conditions. The Context-aware Risk Index (CRI) introduces runtime safety awareness by capturing directional and contextual risk, offering guidance that can inform control behavior without embedding safety logic directly into applications. Together, these mechanisms contribute to a more coordinated and predictable execution environment for autonomous driving.

To support the vehicle’s role as a computing and data service platform, DAVOS integrates Autonomous Vehicle Storage (AVS), which provides hierarchical and queryable storage for continuous multimodal data logging and retrieval across the vehicle lifecycle. Privacy-aware Confidential Computing (PaCC) complements this design by enabling protected execution, purpose-bound access control, and privacy-preserving data processing, allowing vehicle data to be used by services and third-party applications under controlled conditions. The Vehicle Programming Interface (VPI) provides standardized access to data, computation, and system services, reducing coupling between applications and underlying vehicle platforms.

By integrating real-time autonomy support with data-centric vehicle computing, DAVOS moves beyond loosely coordinated subsystems toward a more cohesive and programmable vehicle operating system. This unified foundation improves coordination between driving and data workloads while enabling secure, efficient, and extensible vehicle services. As vehicles continue to evolve toward connected and data-rich platforms, DAVOS provides an architectural path toward scalable vehicle computing, cooperative applications, and future intelligent transportation systems.

\textbf{Availability}
For additional details, system updates, and ongoing developments of the DAVOS project, please visit \url{https://davos4av.org/}



\bibliographystyle{unsrt}
\bibliography{references}

@article{lu2023vehicle,
  title={Vehicle computing: Vision and challenges},
  author={Lu, Sidi and Shi, Weisong},
  journal={Journal of Information and Intelligence},
  volume={1},
  number={1},
  pages={23--35},
  year={2023},
  publisher={Elsevier}
}

@techreport{VectorVehicleOS2022,
  author       = {Benjamin Ram{\'e}l and Marc Weber},
  title        = {Vehicle OS Enabling the Software Defined Vehicle},
  institution  = {Vector Informatik GmbH},
  year         = {2022},
  number       = {Ing{\'e}nieurs de l’Auto No. 881},
  type         = {Technical Article},
  url          = {https://cdn.vector.com/cms/content/know-how/_technical-articles/Vehicle_OS_Ingenieurs_de_l_Auto_PressArticle_202302_EN.pdf},
  note         = {Accessed December 2025}
}

@misc{fordSYNC,
  title        = {Ford {SYNC}: In-Vehicle Connectivity and Infotainment Platform},
  author       = {{Ford Motor Company}},
  howpublished = {\url{https://www.ford.com/technology/sync/}},
  year         = {2024},
  note         = {Accessed: 2025-01-18}
}

@misc{androidAutomotiveOS,
  title        = {Android Automotive {OS}},
  author       = {{Google}},
  howpublished = {\url{https://source.android.com/docs/automotive}},
  year         = {2024},
  note         = {Accessed: 2025-01-18}
}

@misc{qnxRTOS,
  title        = {{QNX} Real-Time Operating System},
  author       = {{BlackBerry QNX}},
  howpublished = {\url{https://blackberry.qnx.com/en/products/qnx-os-for-safety}},
  year         = {2024},
  note         = {Accessed: 2025-01-18}
}

@article{liu2019edge,
  title={Edge computing for autonomous driving: Opportunities and challenges},
  author={Liu, Shaoshan and Liu, Liangkai and Tang, Jie and Yu, Bo and Wang, Yifan and Shi, Weisong},
  journal={Proceedings of the IEEE},
  volume={107},
  number={8},
  pages={1697--1716},
  year={2019},
  publisher={IEEE}
}

@article{shalev2017formal,
  title={On a formal model of safe and scalable self-driving cars},
  author={Shalev-Shwartz, Shai and Shammah, Shaked and Shashua, Amnon},
  journal={arXiv preprint arXiv:1708.06374},
  year={2017}
}

@article{liu2020computing,
  title={Computing systems for autonomous driving: State of the art and challenges},
  author={Liu, Liangkai and Lu, Sidi and Zhong, Ren and Wu, Baofu and Yao, Yongtao and Zhang, Qingyang and Shi, Weisong},
  journal={IEEE Internet of Things Journal},
  volume={8},
  number={8},
  pages={6469--6486},
  year={2020},
  publisher={IEEE}
}

@article{bathla2022autonomous,
  title={Autonomous vehicles and intelligent automation: Applications, challenges, and opportunities},
  author={Bathla, Gourav and Bhadane, Kishor and Singh, Rahul Kumar and Kumar, Rajneesh and Aluvalu, Rajanikanth and Krishnamurthi, Rajalakshmi and Kumar, Adarsh and Thakur, RN and Basheer, Shakila},
  journal={Mobile Information Systems},
  volume={2022},
  number={1},
  pages={7632892},
  year={2022},
  publisher={Wiley Online Library}
}

@article{wang2020safety,
  title={Safety of autonomous vehicles},
  author={Wang, Jun and Zhang, Li and Huang, Yanjun and Zhao, Jian},
  journal={Journal of advanced transportation},
  volume={2020},
  number={1},
  pages={8867757},
  year={2020},
  publisher={Wiley Online Library}
}

@misc{autosar,
  title        = {{AUTOSAR}: Automotive Open System Architecture},
  author       = {{AUTOSAR Partnership}},
  howpublished = {\url{https://www.autosar.org}},
  year         = {2024},
  note         = {Accessed: 2025-01-18}
}

@article{macenski2022robot,
  title={Robot operating system 2: Design, architecture, and uses in the wild},
  author={Macenski, Steven and Foote, Tully and Gerkey, Brian and Lalancette, Chris and Woodall, William},
  journal={Science robotics},
  volume={7},
  number={66},
  pages={eabm6074},
  year={2022},
  publisher={American Association for the Advancement of Science}
}

@inproceedings{kato2018autoware,
  title={Autoware on board: Enabling autonomous vehicles with embedded systems},
  author={Kato, Shinpei and Tokunaga, Shota and Maruyama, Yuya and Maeda, Seiya and Hirabayashi, Manato and Kitsukawa, Yuki and Monrroy, Abraham and Ando, Tomohito and Fujii, Yusuke and Azumi, Takuya},
  booktitle={2018 ACM/IEEE 9th International Conference on Cyber-Physical Systems (ICCPS)},
  pages={287--296},
  year={2018},
  organization={IEEE}
}

@misc{apollo,
  title        = {Apollo: An Open Autonomous Driving Platform},
  author       = {{Baidu Apollo}},
  howpublished = {\url{https://apollo.auto}},
  year         = {2024},
  note         = {Accessed: 2025-01-18}
}

@misc{nvidiaDriveOS,
  title        = {{NVIDIA Drive OS}},
  author       = {{NVIDIA Corporation}},
  howpublished = {\url{https://developer.nvidia.com/drive/drive-os}},
  year         = {2024},
  note         = {Accessed: 2025-01-18}
}

@inproceedings{kronauer2021latency,
  title={Latency analysis of ros2 multi-node systems},
  author={Kronauer, Tobias and Pohlmann, Joshwa and Matth{\'e}, Maximilian and Smejkal, Till and Fettweis, Gerhard},
  booktitle={2021 IEEE international conference on multisensor fusion and integration for intelligent systems (MFI)},
  pages={1--7},
  year={2021},
  organization={IEEE}
}

@article{liu20214c,
  title={4C: A computation, communication, and control co-design framework for CAVs},
  author={Liu, Liangkai and Liu, Shaoshan and Shi, Weisong},
  journal={IEEE Wireless Communications},
  volume={28},
  number={4},
  pages={42--48},
  year={2021},
  publisher={IEEE}
}

@inproceedings{maruyama2016exploring,
  title={Exploring the performance of ROS2},
  author={Maruyama, Yuya and Kato, Shinpei and Azumi, Takuya},
  booktitle={Proceedings of the 13th international conference on embedded software},
  pages={1--10},
  year={2016}
}

@misc{mobaiyen2022systematic,
  title={Systematic Gap Analysis of Robot Operating System (ROS 2) in Real-time Systems},
  author={Mobaiyen, Sahar},
  year={2022}
}

@inproceedings{puck2021performance,
  title={Performance evaluation of real-time ros2 robotic control in a time-synchronized distributed network},
  author={Puck, Lennart and Keller, Philip and Schnell, Tristan and Plasberg, Carsten and Tanev, Atanas and Heppner, Georg and Roennau, Arne and Dillmann, R{\"u}diger},
  booktitle={2021 IEEE 17th International Conference on Automation Science and Engineering (CASE)},
  pages={1670--1676},
  year={2021},
  organization={IEEE}
}

@inproceedings{teper2022end,
  title={End-to-end timing analysis in ROS2},
  author={Teper, Harun and G{\"u}nzel, Mario and Ueter, Niklas and von der Br{\"u}ggen, Georg and Chen, Jian-Jia},
  booktitle={2022 IEEE Real-Time Systems Symposium (RTSS)},
  pages={53--65},
  year={2022},
  organization={IEEE}
}

@article{kouril2024performance,
  title={Performance evaluation of a ROS2 based Automated Driving System},
  author={Kouril, Jorin and Sch{\"a}ufele, Bernd and Radusch, Ilja and Schnor, Bettina},
  journal={arXiv preprint arXiv:2411.11607},
  year={2024}
}

@article{yu2021dynamic,
  title={Dynamic driving environment complexity quantification method and its verification},
  author={Yu, Rongjie and Zheng, Yin and Qu, Xiaobo},
  journal={Transportation Research Part C: Emerging Technologies},
  volume={127},
  pages={103051},
  year={2021},
  publisher={Elsevier}
}

@inproceedings{liu2019towards,
  title={Towards complexity level classification of driving scenarios using environmental information},
  author={Liu, Yongkang and Hansen, John HL},
  booktitle={2019 IEEE Intelligent Transportation Systems Conference (ITSC)},
  pages={810--815},
  year={2019},
  organization={IEEE}
}

@article{kutela2025influence,
  title={The influence of roadway characteristics and built environment on the extent of over-speeding: An exploration using mobile automated traffic camera data},
  author={Kutela, Boniphace and Ngeni, Frank and Ruseruka, Cuthbert and Chengula, Tumlumbe Juliana and Novat, Norris and Shita, Hellen and Kinero, Abdallah},
  journal={International Journal of Transportation Science and Technology},
  volume={17},
  pages={120--130},
  year={2025},
  publisher={Elsevier}
}

@inproceedings{naveiro2024adversarial,
  title={Adversarial Risk Analysis for Automated Lane-Changing in Heterogeneous Traffic},
  author={Naveiro, Roi and R{\'\i}os Insua, David and Caballero, William N},
  booktitle={International Conference on Algorithmic Decision Theory},
  pages={128--143},
  year={2024},
  organization={Springer}
}

@inproceedings{bernhard2022risk,
  title={Risk-based safety envelopes for autonomous vehicles under perception uncertainty},
  author={Bernhard, Julian and Hart, Patrick and Sahu, Amit and Sch{\"o}ller, Christoph and Cancimance, Michell Guzman},
  booktitle={2022 IEEE Intelligent Vehicles Symposium (IV)},
  pages={104--111},
  year={2022},
  organization={IEEE}
}

@inproceedings{wang2024quantitative,
  title={Quantitative analysis of storage requirement for autonomous vehicles},
  author={Wang, Yuxin and He, Yuankai and Wang, Ruijun and Shi, Weisong},
  booktitle={Proceedings of the 16th ACM Workshop on Hot Topics in Storage and File Systems},
  pages={71--78},
  year={2024}
}

@article{xu2024rosfs,
  title={ROSfs: A User-Level File System for ROS},
  author={Xu, Zijun and Wen, Xuanjun and Song, Yanjie and Yin, Shu},
  journal={arXiv preprint arXiv:2406.10635},
  year={2024}
}

@inproceedings{wang2020hydraspace,
  title={HydraSpace: Computational data storage for autonomous vehicles},
  author={Wang, Ruijun and Liu, Liangkai and Shi, Weisong},
  booktitle={2020 IEEE 6th International Conference on Collaboration and Internet Computing (CIC)},
  pages={70--77},
  year={2020},
  organization={IEEE}
}

@article{cai2016iot,
  title={IoT-based big data storage systems in cloud computing: perspectives and challenges},
  author={Cai, Hongming and Xu, Boyi and Jiang, Lihong and Vasilakos, Athanasios V},
  journal={IEEE Internet of Things Journal},
  volume={4},
  number={1},
  pages={75--87},
  year={2016},
  publisher={IEEE}
}

@article{meruje2024databases,
  title={Databases in edge and fog environments: A survey},
  author={Meruje Ferreira, Lu{\'\i}s Manuel and Coelho, Fabio and Pereira, Jos{\'e}},
  journal={ACM Computing Surveys},
  volume={56},
  number={11},
  pages={1--40},
  year={2024},
  publisher={ACM New York, NY}
}

@misc{spyne2025,
  author       = {{Spyne}},
  title        = {Spyne: AI-Powered Conversations and Visuals for Modern Dealerships},
  howpublished = {\url{https://www.spyne.ai/}},
  year         = {2025},
  note         = {Accessed: 2025-12-01}
}

@misc{celantur2025,
  author       = {{Celantur}},
  title        = {Celantur: AI-Powered Image and Video Anonymization Software},
  howpublished = {\url{https://www.spyne.ai/}},
  year         = {2025},
  note         = {Accessed: 2025-12-01}
}

@misc{consumerreports2025ubi,
  author       = {{Consumer Reports}},
  title        = {Usage-Based Car Insurance Can Save You Money, but It Puts Your Data Privacy at Risk},
  howpublished = {\url{https://www.consumerreports.org/money/car-insurance/car-insurance-telematics-pros-and-cons-a5869096072/}},
  year         = {2025},
  note         = {Accessed: 2025-12-01}
}

@misc{geotab2025ubi,
  author       = {{Geotab}},
  title        = {What Is Usage-Based Insurance (UBI)? UBI and Telematics},
  howpublished = {\url{https://www.geotab.com/blog/usage-based-insurance/}},
  year         = {2025},
  note         = {Accessed: 2025-12-01}
}

@misc{geotab2025fleet,
  author       = {{Geotab}},
  title        = {Fleet Management Software and System},
  howpublished = {\url{https://www.geotab.com/fleet-management-software/}},
  year         = {2025},
  note         = {Accessed: 2025-12-01}
}

@misc{dxc2023anonymization,
  author       = {{DXC Technology}},
  title        = {How Anonymization Can Solve Autonomous Driving Data Privacy Challenges},
  howpublished = {\url{https://dxc.com/us/en/insights/perspectives/paper/how-anonymization-can-solve-autonomous-driving-data-privacy-challenges}},
  year         = {2023},
  note         = {Accessed: 2025-12-01}
}

@misc{gallio2023avprivacy,
  author       = {{Gallio}},
  title        = {Data Privacy in Autonomous Vehicles},
  howpublished = {\url{https://gallio.pro/blog/data-privacy-in-autonomous-vehicles/}},
  year         = {2023},
  note         = {Accessed: 2025-12-01}
}

@misc{forrester2024carlistening,
  author       = {Forrester Research},
  title        = {Your Car Is Listening To You—And So Are Hackers},
  howpublished = {\url{https://www.forrester.com/blogs/your-car-is-listening-to-you-and-so-are-hackers/}},
  year         = {2024},
  note         = {Accessed: 2025-12-01}
}

@article{pmc2024drivermonitoring,
  author       = {Various Authors},
  title        = {Technologies for Detecting and Monitoring Drivers' States: A Systematic Review},
  journal      = {PMC (PubMed Central)},
  year         = {2024},
  url          = {https://pmc.ncbi.nlm.nih.gov/articles/PMC11541693/},
  note         = {Accessed: 2025-12-01}
}

@article{liu2022fatigue,
  author       = {Liu, C. and others},
  title        = {A Review of Driver Fatigue Detection and Its Advances on the Use of RGB-D Camera and Deep Learning},
  journal      = {Engineering Applications of Artificial Intelligence},
  year         = {2022},
  url          = {https://www.sciencedirect.com/science/article/abs/pii/S0952197622003967},
  note         = {Accessed: 2025-12-01}
}

@article{pmc2025biosensing,
  author       = {Various Authors},
  title        = {A Comprehensive Review of Unobtrusive Biosensing in Intelligent Vehicles: Sensors, Algorithms, and Integration Challenges},
  journal      = {PMC (PubMed Central)},
  year         = {2025},
  url          = {https://pmc.ncbi.nlm.nih.gov/articles/PMC12189504/},
  note         = {Accessed: 2025-12-01}
}

@misc{insidegnss2018privacy,
  author       = {{Inside GNSS}},
  title        = {Location Privacy Challenges and Solutions},
  howpublished = {\url{https://insidegnss.com/location-privacy-challenges-and-solutions/}},
  year         = {2018},
  note         = {Accessed: 2025-12-01}
}

@article{nyulawreview2023autodata,
  author       = {Various Authors},
  title        = {If Wheels Could Talk: Fourth Amendment Protections Against Auto Data Searches},
  journal      = {NYU Law Review},
  volume       = {98},
  pages        = {2232},
  year         = {2023},
  url          = {https://www.nyulawreview.org/wp-content/uploads/2023/12/98-NYU-L-Rev-2232.pdf}
}

@misc{gizmodo2021onstarice,
  author       = {{Gizmodo}},
  title        = {ICE Is Reportedly Using OnStar Location Data to Track Suspects},
  howpublished = {\url{https://gizmodo.com/ice-is-reportedly-using-onstar-location-data-to-track-s-1846598616}},
  year         = {2021},
  note         = {Accessed: 2025-12-01}
}

@misc{eff2023avprivacy,
  author       = {{Electronic Frontier Foundation}},
  title        = {The Impending Privacy Threat of Self-Driving Cars},
  howpublished = {\url{https://www.eff.org/deeplinks/2023/08/impending-privacy-threat-self-driving-cars}},
  year         = {2023},
  note         = {Accessed: 2025-12-01}
}

@inproceedings{pese2023pricar,
  author       = {Pes\'{e}, M. D. and others},
  title        = {{PRICAR}: Privacy Framework for Vehicular Data Sharing with Third Parties},
  booktitle    = {IEEE Secure Development Conference (SecDev)},
  year         = {2023},
  url          = {https://rtcl.eecs.umich.edu/rtclweb/assets/publications/2023/secdev23-pese.pdf},
  note         = {Discusses indefinite storage of raw sensor data at third parties and privacy risks}
}

@inproceedings{gentry2009fully,
  title={Fully homomorphic encryption using ideal lattices},
  author={Gentry, Craig},
  booktitle={Proceedings of the forty-first annual ACM symposium on Theory of computing},
  pages={169--178},
  year={2009}
}

@inproceedings{dwork2006differential,
  title={Differential privacy},
  author={Dwork, Cynthia},
  booktitle={International colloquium on automata, languages, and programming},
  pages={1--12},
  year={2006},
  organization={Springer}
}

@inproceedings{mayoral2022sros2,
  title={Sros2: Usable cyber security tools for ros 2},
  author={Mayoral-Vilches, Victor and White, Ruffin and Caiazza, Gianluca and Arguedas, Mikael},
  booktitle={2022 IEEE/RSJ International Conference on Intelligent Robots and Systems (IROS)},
  pages={11253--11259},
  year={2022},
  organization={IEEE}
}

@misc{ISO21434,
  author = {{ISO/SAE}},
  title = {ISO/SAE 21434:2021 Road vehicles — Cybersecurity engineering},
  year = {2021},
  url = {https://www.iso.org/standard/70918.html}
}

@inproceedings{casini2019response,
  title={Response-time analysis of ROS 2 processing chains under reservation-based scheduling},
  author={Casini, Daniel and Bla{\ss}, Tobias and L{\"u}tkebohle, Ingo and Brandenburg, Bj{\"o}rn},
  booktitle={31st Euromicro Conference on Real-Time Systems},
  pages={1--23},
  year={2019},
  organization={Schloss Dagstuhl}
}

@inproceedings{choi2021picas,
  title={PiCAS: New design of priority-driven chain-aware scheduling for ROS2},
  author={Choi, Hyunjong and Xiang, Yecheng and Kim, Hyoseung},
  booktitle={2021 IEEE 27th Real-Time and Embedded Technology and Applications Symposium (RTAS)},
  pages={251--263},
  year={2021},
  organization={IEEE}
}

@article{durr2019end,
  title={End-to-end timing analysis of sporadic cause-effect chains in distributed systems},
  author={D{\"u}rr, Marco and Br{\"u}ggen, Georg Von Der and Chen, Kuan-Hsun and Chen, Jian-Jia},
  journal={ACM Transactions on Embedded Computing Systems (TECS)},
  volume={18},
  number={5s},
  pages={1--24},
  year={2019},
  publisher={ACM New York, NY, USA}
}

@article{wu2024timeliness,
  title={Timeliness in autonomous driving: Hype or reality?},
  author={Wu, Tianze and Shi, Weisong},
  journal={IEEE Internet Computing},
  volume={28},
  number={5},
  pages={75--84},
  year={2024},
  publisher={IEEE}
}

@article{bandur2021making,
  title={Making the case for centralized automotive E/E architectures},
  author={Bandur, Victor and Selim, Gehan and Pantelic, Vera and Lawford, Mark},
  journal={IEEE Transactions on Vehicular Technology},
  volume={70},
  number={2},
  pages={1230--1245},
  year={2021},
  publisher={IEEE}
}

@inproceedings{yu2020building,
  title={Building the computing system for autonomous micromobility vehicles: Design constraints and architectural optimizations},
  author={Yu, Bo and Hu, Wei and Xu, Leimeng and Tang, Jie and Liu, Shaoshan and Zhu, Yuhao},
  booktitle={2020 53rd Annual IEEE/ACM International Symposium on Microarchitecture (MICRO)},
  pages={1067--1081},
  year={2020},
  organization={IEEE}
}

@article{padmaja2023exploration,
  title={Exploration of issues, challenges and latest developments in autonomous cars},
  author={Padmaja, Budi and Moorthy, CH VKNSN and Venkateswarulu, N and Bala, Myneni Madhu},
  journal={Journal of Big Data},
  volume={10},
  number={1},
  pages={61},
  year={2023},
  publisher={Springer}
}

@article{pham2021survey,
  title={A survey on security attacks and defense techniques for connected and autonomous vehicles},
  author={Pham, Minh and Xiong, Kaiqi},
  journal={Computers \& Security},
  volume={109},
  pages={102269},
  year={2021},
  publisher={Elsevier}
}

@article{sun2021survey,
  title={A survey on cyber-security of connected and autonomous vehicles (CAVs)},
  author={Sun, Xiaoqiang and Yu, F Richard and Zhang, Peng},
  journal={IEEE Transactions on Intelligent Transportation Systems},
  volume={23},
  number={7},
  pages={6240--6259},
  year={2021},
  publisher={IEEE}
}

@inproceedings{furst2016autosar,
  title={AUTOSAR for connected and autonomous vehicles: The AUTOSAR adaptive platform},
  author={F{\"u}rst, Simon and Bechter, Markus},
  booktitle={2016 46th annual IEEE/IFIP international conference on Dependable Systems and Networks Workshop (DSN-W)},
  pages={215--217},
  year={2016},
  organization={IEEE}
}

\end{document}